\documentclass[11pt, letter]{article}
\usepackage[utf8]{inputenc}
\usepackage[final]{microtype}
\usepackage[letterpaper, margin=1in]{geometry}

\usepackage[hyphens]{url}
\usepackage[style=numeric,
            sorting=none,
            maxnames=6,
            firstinits=true,
            defernumbers=false]{biblatex}
\bibliography{references}

\usepackage{authblk}

\usepackage[hidelinks]{hyperref}
\usepackage{breakurl}

\usepackage{caption}
\usepackage{subcaption}

\usepackage{textgreek}

\usepackage{pdflscape} 
\usepackage{afterpage}

\usepackage{array,booktabs,longtable,tabularx}

\usepackage{amsmath,amsfonts,amssymb,amsthm}

\usepackage{mathtools}

\DeclarePairedDelimiter\ceil{\lceil}{\rceil}

\usepackage{listings}
\usepackage[ruled,vlined,linesnumbered]{algorithm2e}
\usepackage[inline]{enumitem}

\usepackage{diagbox}

\theoremstyle{definition}
\newtheorem{exmp}{Example}[section]

\usepackage{mathtools}
\DeclarePairedDelimiter{\abs}{\lvert}{\rvert}
\let\ket\relax
\DeclarePairedDelimiter{\ket}{\lvert}{\rangle}
\let\bra\relax
\DeclarePairedDelimiter{\bra}{\langle}{\rvert}
\let\braket\relax
\DeclarePairedDelimiterX{\braket}[2]{\langle}{\rangle}{#1 \delimsize\vert #2}

\newcommand{\npat}{r}
\newcommand{\h}{\textsc{h}}
\newcommand{\XOR}{\textsc{cnot}}
\newcommand{\NOT}{\textsc{x}}
\newcommand{\CS}{\textsc{cs}}
\newcommand{\mcx}[1]{\textsc{c}\textsuperscript{#1}\textsc{not}}
\newcommand{\qk}[1]{\texttt{#1}}

\newcommand{\attrspace}{\ceil[\big]{\log_2(a)}}

\begin{document}
\title{EP-PQM: Efficient Parametric Probabilistic Quantum Memory \\ with Fewer Qubits and Gates}

\author[1]{Mushahid Khan}
\author[2]{Jean Paul Latyr Faye}
\author[2]{Udson C. Mendes}
\author[1]{Andriy Miranskyy}

\affil[1]{Department of Computer Science, Ryerson University, Toronto, ON, M5B 2K3, Canada}
\affil[2]{CMC Microsystems, Sherbrooke, QC, J1K 1B8, Canada}
\affil[\space]{mushahid.khan@ryerson.ca, jean.paul.latyr.faye@cmc.ca, udson.mendes@cmc.ca, avm@ryerson.ca}

\date{}
\maketitle

\begin{abstract}
Machine learning (ML) classification tasks can be carried out on a quantum computer (QC) using Probabilistic Quantum Memory (PQM) and its extension, Parametric PQM (P-PQM) by calculating the Hamming distance between an input pattern and a database of $r$ patterns containing $z$ features with $a$ distinct attributes. 
 
For PQM and P-PQM to correctly compute the Hamming distance, the feature must be encoded using one-hot encoding, which is memory-intensive for multi-attribute datasets with $a>2$. We can represent multi-attribute data more compactly by replacing one-hot encoding with label encoding; both encodings yield the same Hamming distance. Implementing this replacement on a classical computer is trivial.
However, replacing these encoding schemes on a QC is not straightforward because PQM and P-PQM operate at the bit level, rather than at the feature level (a feature is represented by a binary string of 0's and 1's).
 
We present an enhanced P-PQM, called EP-PQM, that allows label encoding of data stored in a PQM data structure and reduces the circuit depth of the data storage and retrieval procedures.
We show implementations for an ideal QC and a noisy intermediate-scale quantum (NISQ) device. 
 
Our complexity analysis shows that the EP-PQM approach requires $O\left(z \log_2(a)\right)$ qubits as opposed to $O(za)$ qubits for P-PQM. EP-PQM also requires fewer gates, reducing gate count from $O\left(rza\right)$ to $O\left(rz\log_2(a)\right)$.
 
For five datasets, we demonstrate that training an ML classification model using EP-PQM requires 48\% to 77\% fewer qubits than P-PQM for datasets with $a>2$. 
EP-PQM reduces circuit depth in the range of 60\% to 96\%, depending on the dataset. The depth decreases further with a decomposed circuit, ranging between 94\% and 99\%.

EP-PQM requires less space; thus, it can train on and classify larger datasets than previous PQM implementations on NISQ devices.
Furthermore, reducing the number of gates speeds up the classification and reduces the noise associated with deep quantum circuits.
Thus, EP-PQM brings us closer to scalable ML on a NISQ device.
\end{abstract}

\maketitle

\section{Introduction}\label{sec:intro}
Nowadays, classical algorithms play an important role in finding information in data~\cite{zhou2017machine}. However, due to the large amount of data that we have today, some problems require a lot of resources to be solved~\cite{zhou2017machine}. Recently, quantum computing has emerged as a promising candidate for solving large-scale data problems. Quantum computers (QCs) use quantum mechanical properties, such as superposition and entanglement, to perform computations. Machine learning (ML) is a promising field where QCs, specifically near-term intermediate scale (NISQ) devices, can have potential applications. The purpose of quantum ML is to build quantum-enhanced ML models~\cite{schuld2015introduction,biamonte2017quantum}.
These algorithms have proved to be faster than classical algorithms for a variety of tasks, including supervised and unsupervised ML \cite{russell2016artificial, yom2004advanced, liu2021rigorous}, reinforcement learning \cite{4579244}, and support vector machine~\cite{rebentrost2014quantum}. For review on  quantum ML, see Refs.~\cite{mishra2021quantum, li2021introduction}.

In this paper, we propose an enhancement to a quantum ML model that belongs to the quantum associative memory family of models~\cite{ventura2000quantum,pqmct, trugenbergercarlofull, sousa2020parametric}. Specifically, we focus on Probabilistic Quantum Memory (PQM)~\cite{pqmct, trugenbergercarlofull}, which computes the Hamming distance, deemed $D$, between an $n$-bit input pattern and a database of $r$ patterns of length $n$.

PQM has strong characteristics, such as storing and simultaneously analyzing $r$ patterns while using only $n$ qubits. In other words, on a QC one needs $O(n)$ qubits as opposed to $O(rn)$ bits of associative memory on a classical computer.

Despite this, PQM is not perfect. For example, suppose we have multiple input patterns to compare to the database of patterns. The comparison is made one input string at a time. This means that we need to reload\footnote{This concern may be alleviated by probabilistic cloners for QC~\cite{trugenbergercarlofull}, which may create approximate copies of the states for QC. There are currently no practical probabilistic cloners for QC, but research is ongoing in this area~\cite{PhysRevA.99.012324,PhysRevLett.125.210502,yang2021experimental}.} the database into a QC after every comparison because measuring $D$ collapses the state of the QC~\cite{brun2003comment, schuld2014quantum}. Since the reloading cost is $O(rn)$, it may outweigh the benefits of PQM, making it more efficient to compute $D$ on a classical computer.

PQM can be useful for specialized tasks; e.g., one PQM execution may be sufficient for evaluating the artificial neural network architectures probabilistically without having to initialize the weights~\cite{dos2018quantum}.

PQM can be used in supervised machine learning for pattern classification tasks.  This is done by computing the probability that an input pattern belongs to a given database of patterns~\cite{dossantos2019wnn, sousa2020parametric}.  A variety of datasets were used to demonstrate the applicability of this approach, ranging from detecting breast cancer to finding winning strategies in Tic-Tac-Toe~\cite{dossantos2019wnn, sousa2020parametric}. 

However, the current approach for codifying patterns as binary strings may be inefficient. Let us look at two examples.

\begin{exmp}\label{ex:binary}
Suppose that each of the pattern's features have only two attributes. In this case, we can efficiently codify the value of a feature using a single bit per feature. 
\end{exmp}

\begin{exmp}\label{ex:multi}
What if we have more than two attributes? Let us use $a$ to represent the number of distinct attributes. To codify attributes, we can either use one-hot encoding or label encoding. In one-hot encoding, a single feature is represented by $a$ binary variables, while label encoding maps the attributes to $a$ integers. 
Therefore, one-hot encoding requires $a$ bits to represent a feature, while label encoding only needs $\attrspace$ bits, where  $\ceil{\cdot}$ denotes ceiling function. Thus, label encoding is more memory-efficient than one-hot encoding. This difference is significant as real-world dataset (e.g. NSL-KDD~\cite{tavallaee2009kdd}) may contain hundreds of distinct attributes. For example, one-hot encoding requires 100 bits for a feature with 100 distinct attributes, whereas label encoding requires only 7 bits. This exponential decrease in the number of qubits and gates may make NISQ devices capable of processing complex multi-attribute patterns.

Consider patterns containing three features: an input pattern $\mu_0 =$~``\texttt{A~A~A}'' and a database with two patterns $\mu_1$ = ``\texttt{B B B}'' and $\mu_2$ = ``\texttt{C C A}''. We would like to compute the distance between $\mu_0$ and two patterns in the database. It is obvious that $D(\mu_0, \mu_1) = 3$ and $D(\mu_0, \mu_2) = 2$.

Let us now demonstrate how the above example is implemented using PQM. As PQM operates at the bit level, the patterns must be converted to bit strings. We will use the label encoding to represent the three attributes, which requires two bits per attribute. Suppose that ``\texttt{A}'' is codified as ``\texttt{00}'', ``\texttt{B}'' as ``\texttt{01}'', and ``\texttt{C}'' as ``\texttt{11}''. This encoding converts the patterns to the following bit strings: $\mu_0 =$~``\texttt{00~00~00}'', $\mu_1$ = ``\texttt{01 01 01}'', and $\mu_2$ = ``\texttt{11 11 00}''. PQM operates on individual bits, thus $D_{\textrm{label}}(\mu_0, \mu_1) = 3$ and $D_{\textrm{label}}(\mu_0, \mu_2) = 4$.

As we can see, $D(\mu_0, \mu_1) > D(\mu_0, \mu_2)$, but $D_{\textrm{label}}(\mu_0, \mu_1) <  D_{\textrm{label}}(\mu_0, \mu_2)$. Consequently, label encoding, while efficiently utilizing memory, may lead to erroneous results.

To overcome this problem, we must resort to one-hot encoding, in which attributes are codified as follows: ``\texttt{A}''  as ``\texttt{100}'', ``\texttt{B}'' as ``\texttt{010}'', and ``\texttt{C}'' as ``\texttt{001}''. The patterns will now be represented by $\mu_0 =$~``\texttt{100~100~100}'', $\mu_1$ = ``\texttt{010 010 010}'', and $\mu_2$ = ``\texttt{001 001 100}''. PQM will yield $D_{\textrm{one-hot}}(\mu_0, \mu_1) = 6$ and $D_{\textrm{one-hot}}(\mu_0, \mu_2) = 4$. Note that $D_{\textrm{one-hot}} = 2 D$ (see Appendix~\ref{sec:one-hot-and-d} for details); thus, one-hot encoding yields the desired values of $D$.

Therefore, one-hot encoding allows PQM to compute $D$ correctly, whereas label encoding does not. However, one-hot encoding consumes more space: nine qubits will be required instead of six.
\end{exmp}

Based on the above examples, to compute $D$ correctly, it appears that PQM can only use one-hot encoding for features with multiple attributes. Consequently, the pattern must be represented using a large amount of space. Modern QCs, however, have a small number of qubits and limited coherence. Therefore, we need to find a way to use label encoding to reduce space constraints and utilize qubits more efficiently. 

In this paper, our contribution is as follows. We generalize the PQM algorithm to compute the Hamming distance $D$ for features codified with label encoding. Specifically, we improve the parametric PQM (P-PQM) classification algorithm designed for modern noisy intermediate-scale quantum (NISQ) computers~\cite{sousa2020parametric} to use fewer qubits and quantum gates. This efficient version of P-PQM, which we call EP-PQM, results in 
\begin{enumerate}
    \item Space savings, as for $z$ features the number of required qubits reduces from $O(za)$ to $O\left(z \log_2(a)\right)$ for $a>2$;
    \item Improved computational efficiency, as the amount of quantum gates needed to execute the PQM algorithm is proportional to the number of bits required to represent a pattern. The number of gates reduces from  $O\left(rza\right)$ to $O\left(rz\log_2(a)\right)$.
\end{enumerate}
With EP-PQM requiring fewer qubits and gates to implement PQM than P-PQM, it becomes a better candidate for applications in NISQ devices.

The rest of the paper is structured as follows. Section~\ref{sec:pqm} recaps the PQM and P-PQM algorithms~\cite{sousa2020parametric}. In Section~\ref{sec:eppqm}, we introduce the EP-PQM. Section~\ref{sec:exp} provides sample experiments with ML datasets. Finally, in Section~\ref{sec:summary} we present our final remarks and conclusions.

\section{Probabilistic Quantum Memory (PQM)}\label{sec:pqm}
In this section, we review the PQM and P-PQM algorithms. 
The PQM~\cite{pqmct, trugenbergercarlofull} algorithm introduces a data structure that allows computing the $D$ distance at the bit level between a binary input pattern and all other binary patterns stored in a superposition state. 
PQM has two parts: storing information and retrieving information, discussed in Sections~\ref{sec:orig_storage} and~\ref{sec:orig_retrieval}, respectively. The core quantum computing fundamentals needed to implement these algorithms are introduced in Appendix~\ref{sec:qc}. 

To store information, PQM will take a dataset of $r$ binary patterns, each of $n$ bits and store them in a superposition state with equal probability. Given an input, to retrieve information, the memory quantum state is rotated within the subspace defined by the stored patterns. 
The resulting amplitudes are peaked on the stored patterns which are closest in Hamming distance $D$ to the input. 

P-PQM is similar in nature to PQM, but adds an extra parameter that may improve ML classification.

\subsection{PQM and P-PQM: Storing Information} \label{sec:orig_storage}
This section recaps the existing PQM and P-PQM data storage processes, leveraging the same algorithm. 

Formally, the storing information part of the algorithm receives a dataset of $r$ binary patterns, each containing $n$ bits: $data = \cup_{i=1}^{r}{ \{ p^i \} }$. To store the patterns, three quantum registers are needed: the input register $p$, the memory register $m$, and the auxiliary two-qubit register $u$. The input register state $\ket{p}$ will hold every pattern of length $n$. $\ket{m}$ is the memory register which will store each pattern $p^i$ by the end of the algorithm. The auxiliary two-qubit register state $\ket{u}$ is used to keep tabs on which patterns are stored in memory and which ones need to be processed. The first qubit in  $\ket{u}$ is used to change the second qubit in  $\ket{u}$. The second qubit in $\ket{u}$ indicates whether a pattern has been already stored or not. In this case, $\ket{u}$ = 1 indicates that the pattern has not been stored yet. To make a copy of the $n$ bits from the pattern in  $\ket{p}$ to the respective  register $\ket{m}$, the algorithm checks if the pattern has been stored on  $\ket{m}$ by checking if second qubit in $\ket{u}$ is $1$. If it is $1$, the pattern will be copied to $\ket{m}$. 
All three quantum registers are initialized in the state $\ket{0}$ and the algorithm initial state is
\begin{equation}\label{eq:pqm_storage_register}
\ket{\psi_0}_1 =  \ket{ \underbrace{0_1 0_2 \ldots 0_n}_{\ket{p}}; \underbrace{01}_{\ket{u}}; \underbrace{0_1 0_2 \ldots 0_n}_{\ket{m}}}.
\end{equation}

The storage process loads each pattern (represented as a binary string) into $\ket{p}$ and then stores them in $\ket{m}$ in a superposition with equal probabilities. Once the patterns are loaded into $\ket{m}$, they will be processed to find the closest pattern match as we will discuss in Section~\ref{sec:orig_retrieval}.

The storage process is given in Algorithm~\ref{alg:pqm_storing}, see Ref.~\cite{sousa2020parametric} for an in-depth explanation of the storing algorithm. This algorithm uses common quantum computing gates recapped in Appendix~\ref{sec:qc}. In addition to these gates, Step 7 of the algorithm uses a two-qubit control gate $\CS^j$, which adds a pattern $p^i$ to memory register with uniform amplitudes. It is defined as follows:
$$\CS^j = \ket{0}\bra{0} \otimes 1 + \ket{1}\bra{1} \otimes S^j,$$
where $j \in \mathbb{Z}$ and $j=1,2,$\ldots$,r$, and
$$ S^j = \begin{bmatrix}\sqrt{\frac{j-1}{j}}&\frac{1}{j} \\ 
                        \frac{-1}{\sqrt{j}} & \sqrt{\frac{j-1}{j}}\end{bmatrix}.$$
Thus,
$$\CS^j = \begin{bmatrix} 1 & 0 & 0&0\\ 0 & 1 & 0&0\\ 0 & 0 &\sqrt{\frac{j-1}{j}}&\frac{1}{j} \\ 0 & 0 & \frac{-1}{j}&\sqrt{\frac{j-1}{j}}\end{bmatrix}.$$

The final result of the algorithm is the state
\begin{equation}\label{eq:pqm_storage_final_state}
\ket{\psi^r_8}_1 =
\frac{1}{\sqrt{r}} \sum_{k = 1}^{\npat} \ket{0_1 0_2 \ldots 0_n;0 1; m_1^k m_2^k \ldots m_n^k}.
\end{equation}
Note that in the retrieval phase, we are only concerned with the memory-register state $\ket{m}$.

\begin{algorithm}[ht]
    \SetKwInOut{Input}{input}
    \SetKwInOut{Output}{output}
    \Input{The initial state $\ket{\psi_0}_1$  given in Eq.~\eqref{eq:pqm_storage_register}} 
    \Output{Return $\ket{\psi^r_8}_1$ defined in Eq.~\eqref{eq:pqm_storage_final_state}}
    $\ket{\psi_1^1} = \ket{\psi_0}_1$ \\ 
    \ForEach{$p^i \in data$}{ \label{str:loop}
        Load $p^i$ into quantum register $\ket{p}_n$ \\ 
        $\ket{\psi_2^i} = \prod_{j=1}^n 
        \mcx{2}_{p_j^i,u_2,m_j}\ket{\psi_1^i}$ \\ 
        $\ket{\psi_3^i} = \prod_{j=1}^n \NOT_{m_j} 
        \XOR_{p_j^i,m_j}\ket{\psi_2^i}$ \\ 
        $\ket{\psi_4^i} = \mcx{n}_{m_1\cdots m_n,u_1}\ket{\psi_3^i}$ \\ 
        \label{str:step3}
        $\ket{\psi_5^i} = \CS_{u_1,u_2}^{\npat+1-i}\ket{\psi_4^i}$ \\ 
        \label{str:step4}
        $\ket{\psi_6^i} = \mcx{n}_{m_1\cdots m_n,u_1}\ket{\psi_5^i}$ \\ 
        \label{str:step5}
        $\ket{\psi_7^i} = \prod_{j=1}^n \XOR_{p_j^i,m_j} 
        \NOT_{m_j}\ket{\psi_6^i}$ \\
        $\ket{\psi_8^i} = \prod_{j=1}^n 
        \mcx{2}_{p_j^i,u_2,m_j}\ket{\psi_7^i}$ \\
        $\ket{\psi_1^{i+1}} = \ket{\psi_8^i}$ \\
    Unload $p^i$ from quantum register $\ket{p}_n$ 
    }
    The final state after completion of the for-loop will be $\ket{\psi^r_8}_1$
    \caption{PQM and P-PQM storage algorithm~\cite{sousa2020parametric}. In control gate $\mcx{\textalpha}_{m_1\cdots m_n,u_1}$ qubits $m_1\cdots m_n$ act as controls and qubit $u_1$
is a target. For more details, see Appendix~\ref{sec:qc}.}
    \label{alg:pqm_storing}
\end{algorithm}

\subsection{PQM and P-PQM: Retrieving Information}\label{sec:orig_retrieval} 
PQM and P-PQM retrieval processes~--- described in Sections~\ref{sec:pqm_retrieval} and~\ref{sec:ppqm_retrieval}, respectively~--- are similar. PQM retrieval algorithm has only implementation for fault-tolerant QCs, while P-PQM also has an implementation designed for NISQ devices~\cite{sousa2020parametric}. 

\subsubsection{PQM: Retrieving Information}\label{sec:pqm_retrieval}
The algorithm for retrieving information  relies on the memory register $m$ in state $\ket{\psi^r_8}_1$ (i.e., the output of the storage algorithm). This state is then further manipulated to perform pattern analysis as described below.

PQM uses the Hamming distance $D$ between a target pattern and all patterns, which are stored in a superposition, to indicate probabilistically the chances of the target pattern being in the memory. This algorithm uses three quantum registers, namely $t$, $m$, and $c$:
\begin{equation}\label{eq:pqm_retrieve_registry}
     \underbrace{t_1 t_2 \ldots t_n}_{t}; \underbrace{m_1 m_2 \ldots m_n}_{m}; \underbrace{c}_{c}.
\end{equation}
The target pattern, deemed $T$ and represented by bits $\tau_1 \tau_2\ldots \tau_n$, is loaded into register state $\ket{t}$; $\ket{m}$ contains all the stored patterns from the storage algorithm; and $\ket{c}$ contains a control qubit initialized in state $\ket{0}$. Once the input has been loaded on to $\ket{t}$ and the stored patterns are in $\ket{m}$, the full initial quantum state is
\begin{equation}\label{eq:pqm_state_initial}
\ket{\psi_0}_2 = \frac{1}{\sqrt{\npat}} \sum_{k = 1}^{\npat} \ket{t_1 t_2 \ldots t_n;m_1^k m_2^k \ldots m_n^k;0},
\end{equation}
where $r$ is the total number of stored patterns, $t_1 t_2\ldots t_n$ are qubits used to store corresponding bits $\tau_1 \tau_2 \ldots \tau_n$ of the target pattern $T$, and $m_1^k m_2^k \ldots m_n^k$ is the $k$-th stored pattern.
\begin{algorithm}[t]
 \SetKwInOut{Input}{input}
    \SetKwInOut{Output}{output}
    \Input{The initial state $\ket{\psi_0}_2$  given in Eq.~\eqref{eq:pqm_state_initial}} 
    \Output{1) The value of $c$ and 2) one of the stored patterns that has minimum $D$ with the target pattern in a given run}
    $\ket{\psi_1 } = \h_c\ket{\psi_0}_2$ \\
    $\ket{\psi_2 } = \prod_{j=1}^n \NOT_{m_j} \XOR_{t_j, m_j} \ket{\psi_1 }$ \\
    $\ket{\psi_3 } = \prod_{e=1}^n \left(G\tilde{U}^{-2}\right)_{c,m_e} \prod_{j=1}^n \tilde{U}_{m_j} \ket{\psi_2}$ \\
    $\ket{\psi_4 } = \h_c\prod_{j=n}^1  \XOR_{t_j, m_j}\NOT_{m_j} \ket{\psi_3 }$\\
    Measure qubit $\ket{c}$\\
    \If{$c == 0$}{
        Measure the memory to obtain the desired state.
    }
    \caption{PQM retrieval algorithm --- fault-tolerant implementation~\cite{sousa2020parametric}.}
    \label{alg:recover_orig}
\end{algorithm}

The retrieval process is summarized in Algorithm~\ref{alg:recover_orig}. In Step 1, we apply the Hadamard gate on to the control qubit $\ket{c}$ to get
\begin{equation}\label{eq:pqm_state}
\ket{\psi_1} = \frac{1}{\sqrt{2\npat}} \left(\sum_{k = 1}^{\npat} \ket{t_1 t_2 \ldots t_n;m_1^k m_2^k \ldots m_n^k;0} + \sum_{k = 1}^{\npat} \ket{t_1 t_2 \ldots t_n;m_1^k m_2^k \ldots m_n^k;1} \right). 
\end{equation}

Step 2 sets the $j$-th qubit in register $\ket{m}$ to $\ket{1}$ if the $j$-th qubit of $\ket{t}$ and $\ket{m}$ are the same or to $\ket{0}$ if they differ.

Step 3 computes $D$ between the target pattern and all patterns in $\ket{m}$. The number of zeros in $\ket{m}$ (representing the qubits that differ between memory and target string) is computed.
Operator $\tilde{U}$, used in this step, is defined as
\begin{equation}\label{eq:u_pqm}
\tilde{U} = \left[\begin{array}{cc}
\exp\left(\frac{i \pi}{2n}\right) &   0 \\
0                  &   1
\end{array}\right], 
\end{equation}
where $i$ denotes unit imaginary number. First, $\tilde{U}$ is applied to each qubit in  $\ket{m}$. Then $\tilde{U}^{-2}$ (which is $\tilde{U}$ to the power of $-2$) is applied to each qubit in $\ket{m}$ if the qubit is in state $\ket{c} = \ket{1}$. This control operator is denoted by $G$. As per~\cite{pqmct}, $G \tilde{U}^{-2}$ is formally defined as
$$G \tilde{U}^{-2} = \ket{0}\bra{0} \otimes 1 + \ket{1}\bra{1} \otimes \tilde{U}^{-2}.$$

Step 4 reverts register $\ket{m}$ to its original state and the Hadamard gate \textrm{\textsc{h}} is applied to the control qubit in $\ket{c}$; this operation is denoted by $\h_c$. 
After Step 4, the state will be:
\begin{equation}\label{eq:retrieval_state_before_measuring_c}
\ket{\psi_4} = \frac{1}{\sqrt{\npat}}  \left(\sum_{k = 1}^{\npat} \cos\frac{\pi}{2n}d_k\ket{t_1 t_2 \ldots t_n;m_1^k m_2^k \ldots m_n^k;0} \\ + \sum_{k = 1}^{\npat}\sin\frac{\pi}{2n}d_k \ket{t_1 t_2 \ldots t_n;m_1^k m_2^k \ldots m_n^k;1} \right),
\end{equation}
where $d_k$ is $D$ between the target pattern $t$ and stored pattern $m^k$.

Step 5 measures register $\ket{c}$. A target pattern similar to the stored patterns increases the probability of measuring $\ket{c} = \ket{0}$. Otherwise, if the target pattern is dissimilar, then the probability of $\ket{c} = \ket{1}$ increases. If $\ket{c}$ is measured in $\ket{0}$ (Step 6), then measuring the qubits in the memory register (in Step 7) will return the binary pattern from the set of stored patterns that produced the minimum $D$ with the target pattern in a given run of the algorithm, with the following probability: 

\begin{equation}\label{eq:retrieval_after_register_c_is_0}
P \left( m^k \right)=
\begin{cases}
0 & \quad  \text{if}~\ket{c} = \ket{1}\\
\frac{1}{rP(\ket{c} = \ket{0})} \cos^2 \left(\frac{\pi}{2n}d_k \right) & \quad  \text{otherwise} \\
\end{cases}.
\end{equation}
This probability peaks around the patterns which have smallest $D$ to $t$.

\subsubsection{P-PQM: Retrieving Information} \label{sec:ppqm_retrieval}
In a nutshell, P-PQM operates as the PQM, but with the addition of a scale parameter $\nu \in (0,1]$ in the retrieval algorithm. The parameter $\nu$ is used to compute weighted $D$, which may improve performance of the classifier~\cite{sousa2020parametric}. 

To integrate $\nu$ into the retrieval algorithm, Eq.~\eqref{eq:u_pqm} is redefined as
\begin{equation}\label{eq:u_p-pqm}
{U} = \left[\begin{array}{cc}
\exp\left(\frac{i \pi}{2n\nu}\right) &   0 \\
0                  &   1
\end{array}\right]. 
\end{equation}
Note that the quantum circuit depth and complexity are independent of $\nu$, and that P-PQM reduces to PQM when $\nu=1$. Let us give a brief summary of both fault-tolerant and NISQ implementations (see~\cite{sousa2020parametric} for additional details).

\paragraph{Fault-tolerant implementation.}
The algorithm for the quantum implementation of the P-PQM retrieval procedure is almost identical to the PQM one (shown in Algorithm~\ref{alg:recover_orig}). The only difference is that Eq.~\eqref{eq:u_pqm} is replaced with Eq.~\eqref{eq:u_p-pqm}. The input into the algorithm, as in the PQM case, is given by Eq.~\eqref{eq:pqm_state}.

\paragraph{NISQ implementation.} 
The NISQ implementation of the P-PQM retrieval algorithms is shown in Algorithm~\ref{alg:recover_ppqm}.
This implementation requires only two registers, namely, $m$ and $c$:
\begin{equation}\label{eq:ppqm_retreival_registrers}
    \underbrace{m_1 m_2 \ldots m_n}_{m}; \underbrace{c }_{c},
\end{equation}
and the input is stored classically.
Once the stored patterns from the storing part of the algorithm are in $\ket{m}$, the full initial quantum state is:
\begin{equation}\label{eq:ppqm_state}
\begin{split}
\ket{\psi_0}_3 &= \frac{1}{\sqrt{\npat}} \sum_{k = 1}^{\npat} \ket{m_1^k m_2^k \ldots m_n^k;0}.
\end{split}
\end{equation}
Because the input is stored classically, all the control operators from the input register to the memory register are removed from the circuit and replaced with a $\NOT$ operator. Essentially, the input pattern is processed dynamically without the need for a dedicated register.
In the NISQ implementation~\cite{sousa2020parametric}, Steps 2 and 4 of the fault-tolerant implementation (Algorithm~\ref{alg:recover_orig}) are modified. Specifically, they are replaced by Steps 2--4 and 6--8 of Algorithm~\ref{alg:recover_ppqm}. In this NISQ implementation, rather than applying $\NOT$ based on $\XOR$ to each qubit, we examine each $\tau_j$ bit in $T$ on the classical computer and apply $\NOT$ only if $\tau_j=1$. This leads to ``inversion'' of logic; thus, we are now interested in measuring $\ket{m}$ when $c = 1$ rather than when $c = 0$ (compare Steps 6--7 of Algorithm~\ref{alg:recover_orig} with Steps 11--12 of the Algorithm~\ref{alg:recover_ppqm}). 

\begin{algorithm}[t]
\SetKwInOut{Input}{input}
    \SetKwInOut{Output}{output}
    \Input{The initial state $\ket{\psi_0}_3$  given in Eq.~\eqref{eq:ppqm_state}} 
    \Output{1) The value of $c$ and 2) one of the stored patterns that has minimum $D$ with the target pattern in a given run }
    $\ket{\psi_1 } = \h_c\ket{\psi_0}_3$ \\
    \ForEach{$\tau_j \in T$}{  
    \If{$\tau_j == 1$}
    {
    $\ket{\psi_2 } = \NOT_{m_j} \ket{\psi_1 }$ \\
    }
    }
    $\ket{\psi_3 } = \prod_{e=1}^n \left(GU^{-2}\right)_{c,m_e} \prod_{j=1}^n U_{m_j} \ket{\psi_2 }$ \\
    \ForEach{$\tau_j \in T$}{ 
    \If{$\tau_j == 1$} {
    
    $\ket{\psi_4 } = \NOT_{m_j} \ket{\psi_3}$ \\
    }
    }
    $\ket{\psi_5 } = \h_c \ket{\psi_4 }$\\
    Measure qubit $\ket{c}$\\
    \If{$c == 1$}{
        Measure the memory register to obtain the desired state.
    }
    \caption{P-PQM retrieval algorithm --- NISQ implementation~\cite{sousa2020parametric}.}
    \label{alg:recover_ppqm}
\end{algorithm}

\subsection{Post-processing on classical computer}\label{sec:post-processing}
PQM- and P-PQM-based ML have probabilistic nature. 
To get $N$ measurements, we need to run the storage and retrieval algorithms $N$ times (because after the memory registers are measured, we need to re-initialize the database).

We collect $N$ measurements of $\ket{c}$ for PQM-based ML classification (whether PQM or P-PQM). Suppose we measure $M \leq N$ instances when $\ket{c} = 0$ (for fault-tolerant Algorithm~\ref{alg:recover_orig}) or when $\ket{c} = 1$ (for NISQ Algorithm~\ref{alg:recover_ppqm}). Then the affinity of an input pattern belonging to a given database of patterns, deemed $\rho$, is given by $\rho = M/N$. The closer $\rho$ is to $1$ --- the closer the input pattern is to the database of patterns. 

To infer the class/label of an input pattern, an analyst needs to construct individual pattern databases for each class/label. Next, the analyst will compute $\rho$ for each database. Finally, the analyst will assign a label to the input pattern based on the label of the database with the highest $\rho$. In this paper, we do not compute $\rho$ as our goal is to improve the storage and retrieval process used to obtain individual measurements. To learn more about PQM-based ML classification, see Ref.~\cite{sousa2020parametric}.

Note that the last two steps of Algorithms~\ref{alg:recover_orig} and~\ref{alg:recover_ppqm} suggest to measure the values of specific patterns in the database. On an ideal QC, we can get rid of these steps, as all the required information needed to compute $\rho$ would be given to us by measuring qubit $\ket{c}$. However, for NISQ devices, the situation is different: we may end up with measuring a pattern that has not been stored in the database. This happens due to a noisy nature of NISQ devices. In this case, we may need to implement an additional post-processing scheme. For example, one may assume that information about the value of $\ket{c}$ is important and should be included in the computation of $\rho$ independent of the value of the pattern. Another approach would involve discarding the measurements of patterns that have not been present in the database. Answering this question is outside of the scope of this paper.

\section{EP-PQM}\label{sec:eppqm}
In this section, we cover our extension of P-PQM. Storing of information is discussed in Section~\ref{sec:our_storing}, retrieval~--- in Section~\ref{sec:our_retrieval}. For both storage and retrieval, we design fault-tolerant and NISQ implementations. Both of these implementations of EP-PQM typically require fewer qubits and gates than PQM and P-PQM implementations, as we will demonstrate below.

\subsection{EP-PQM: Storing Information}\label{sec:our_storing}
\paragraph{Fault-tolerant implementation.}
For a fault-tolerant implementation, we will reuse the PQM and P-PQM storage procedures shown in Algorithm~\ref{alg:pqm_storing}.

\paragraph{NISQ implementation.} 
EP-PQM storage procedure suitable for NISQ device is given in Algorithm~\ref{alg:ep-pqm_storing}. The latter requires two registers, the memory register $\ket{m}$ with $n$ qubits and an auxiliary two-qubit register $\ket{u}$, as defined in Section~\ref{sec:orig_storage}. The algorithm starts with the following initial state 
\begin{equation}\label{eq:hybrid_storage_register}
\ket{\psi_0}_4 = \ket{  \underbrace{01}_{\ket{u}}; \underbrace{0_1 0_2 \ldots 0_n}_{\ket{m}}},
\end{equation}
where all the qubits in memory register are in state $\ket{0}$ and the auxiliary register is in state $\ket{01}$.
Since in EP-PQM the input patterns do not require a dedicated register, as they can be stored in a classical computer, this algorithm requires less qubits for the storage part. Indeed, comparing Eqs.~\eqref{eq:hybrid_storage_register} and~\eqref{eq:pqm_storage_register}, we see that EP-PQM storage part uses $n+2$ qubits versus $2n+2$ qubits for PQM and P-PQM. After execution of Algorithm~\ref{alg:ep-pqm_storing}, the final state is a superposition of the input patterns with equal probabilities:
\begin{equation}\label{eq:ep_pqm_state_store}
\ket{\psi^r_6}_4= 
\frac{1}{\sqrt{r}} \sum_{k = 1}^{\npat} \ket{0 1; m_1^k m_2^k \ldots m_n^k}.
\end{equation}

\begin{algorithm*}[ht]
    \SetKwInOut{Input}{input}
    \SetKwInOut{Output}{output}
    \Input{The initial state $\ket{\psi_0}_4$ given in Eq.~\eqref{eq:hybrid_storage_register}}  
    \Output{Return $\ket{\psi^r_6}_4$ defined in Eq.~\eqref{eq:ep_pqm_state_store}}
    
    Prepare the initial state $\ket{\psi_1^1} =  \ket{\psi_0}_4$ \\ 
    \ForEach{$p^i \in data$}{ 
    \tcc{$p^i_j$ denotes the $j$-th bit of the $i$-th pattern $p_i$.}
    \ForEach{$p^i_j \in p^i$}{
        \If{$p^i_j == 1$}{$\ket{\psi_2^i} =\XOR_{u_2,m_j}\ket{\psi_1^i}$ }
        \Else{$\ket{\psi_2^i} =\NOT_{m_j} \ket{\psi_1^i}$}
    }
    $\ket{\psi_3^i} = \mcx{n}_{m_1\cdots m_n,u_1}\ket{\psi_2^i}$ \\ 
    $\ket{\psi_4^i} = \CS_{u_1,u_2}^{\npat+1-i}\ket{\psi_3^i}$ \\ 
    $\ket{\psi_5^i} = \mcx{n}_{m_1\cdots m_n,u_1}\ket{\psi_4^i}$ \\ 
    \ForEach{$p^i_j \in p^i$}{
        \If{$p^i_j == 1$}{$\ket{\psi_6^i} =\XOR_{u_2,m_j}\ket{\psi_5^i}$}
        \Else{$\ket{\psi_6^i} =\NOT_{m_j} \ket{\psi_5^i}$}
    }
     $\ket{\psi_1^{i+1}} = \ket{\psi_6^i}$
    }
        The final state after completion of the for-loop will be $\ket{\psi^r_6}_4$
    \caption{EP-PQM storage algorithm --- NISQ implementation.}
    \label{alg:ep-pqm_storing}
\end{algorithm*}

In Algorithm~\ref{alg:ep-pqm_storing}, since the input pattern is stored classically, all the control operators acting on the input register are removed. Instead, we can classically check if the input pattern consists of 0 or 1. As a result, Steps 4, 5, 9, and 10 of Algorithm~\ref{alg:pqm_storing} are changed. These changes can be seen in Steps 3--7, and 11--15 of Algorithm~\ref{alg:ep-pqm_storing}. Steps 3--7 are used to make a copy of the $n$ bits of $p^i$ to the memory register, if the  $u_2$ is flagged as  $1$, and then fill with $1$'s all the bits in the memory register which are equal to the respective bits in $p^i$. Steps 11--15 will reverse the work done in Steps 3--7.

\subsection{EP-PQM: Retrieving Information}\label{sec:our_retrieval}
\paragraph{Fault-tolerant implementation.}

As explained in Section~\ref{sec:post-processing}, after $N$ measurements, PQM outputs the probability of a target pattern being close to patterns in the database at the bit level.
As discussed in Section~\ref{sec:intro}, this approach is ineffective if we compute $D$ for symbols represented by multiple bits. For this reason, we extend the information retrieval part of PQM given in Algorithm~\ref{alg:recover_orig}.

Our extension generalizes the PQM algorithm to compute $D$ for features codified with label encoding. The extension requires four registers, namely, $t$, $m$, $c$, and $h$:
\begin{equation}\label{eq:eppqm_retrieve_registry}
     \underbrace{t_1 t_2 \ldots t_n}_{t}; \underbrace{m_1 m_2 \ldots m_n}_{m}; \underbrace{c}_{c};\underbrace{h_1 h_2 \ldots h_z}_{h}.
\end{equation}

Registers $t$, $m$, and $c$ are the same as in Eq.~\eqref{eq:pqm_retrieve_registry}, i.e., the PQM case discussed in Section~\ref{sec:pqm_retrieval}. 

The register $h$ is used to compare features. The number of qubits in $h$ is equal to the number of features in the pattern, deemed $z$. Note that $z = n/d$, where $d$ is the number of bits required to represent an attribute of a feature, i.e., $d=\attrspace$. The $j$-th qubit in register $h$ is set to $\ket{1}$ if the binary string of length $d$ that represent $j$-th feature of $T$ is the same as the corresponding binary string in $m$. With $\ket{h}$ included, the initial quantum state for retrieval of information is
\begin{equation}\label{eq:our_state}
\begin{split}
\ket{\psi_0}_5 &= \frac{1}{\sqrt{\npat}} \sum_{k = 1}^{\npat} \ket{t_1 t_2 \ldots t_n;m_1^k m_2^k \ldots m_n^k;0;1_1 1_2 \ldots 1_z},
\end{split}
\end{equation}
where all qubits in register $\ket{h}$ are initially in state $\ket{1}$ and the register $\ket{c}$ in state $\ket{0}$. 

With this extension, the retrieval algorithm is changed, as shown in Algorithm~\ref{alg:recover_ours}. The first and second steps are the same as the original  Algorithm~\ref{alg:recover_orig}. In Step 3, results from Step 2 are used to update register $\ket{h}$. Given that each symbol is represented by a binary string of length $d$ and each input is of length $n$, Step 3 will set the $j$-th qubit in $\ket{h}$ to $\ket{0}$. This will happen if the binary string of length $d$, that represent the $j$-th symbol of $T$, is not the same as the corresponding binary string in $\ket{m}$. With register $\ket{h}$ updated in Step 3, $\ket{h}$ can be used in place of $\ket{m}$ for Step 4. In Step 4, we use operator $W$ instead of $U$, which is defined as
\begin{equation*}
W = \left[\begin{array}{cc}
\exp\left(\frac{i \pi}{2z\nu}\right) &   0 \\
0                  &   1
\end{array}\right].
\end{equation*}
We use $W$ to calculate $D$ at the feature level. In Steps 5 and 6, inverse transformations of Steps 2 and 3 are applied and $\h$ gate is applied to the control qubit. In Step 7, $\ket{c}$ is measured. If $\ket{c}$ is measured in state $\ket{0}$, it means that the input is close to all stored patterns in the dataset. This  probability  is  peaked  around  those  patterns  which have the smallest $D$ to the input at the feature level. The highest  probability of retrieval thus occurs for patterns which are most similar to the input at the feature level. Finally, Steps 7--9  are identical to Steps 5--7 of Algorithm~\ref{alg:recover_orig}.

\begin{algorithm*}[t]
\SetKwInOut{Input}{input}
    \SetKwInOut{Output}{output}
    \Input{The initial state $\ket{\psi_0}_5$  given in Eq.~\eqref{eq:our_state}} 
    \Output{1) The value of $c$ and 2) one of the stored patterns that has minimum $D$ with the target pattern in a given run}
    $\ket{\psi_1 } = \h_c\ket{\psi_0}_5$ \\
    $\ket{\psi_2 } = \prod_{j=1}^n \NOT_{m_j} \XOR_{t_j, m_j} \ket{\psi_1 }$ \\
    $\ket{\psi_3 } = \prod_{j=1}^z \NOT_{h_j} \mcx{d}_{m_{d  (j-1) + 1 } m_{d(j-1)+2}\ldots m_{d j}, h_j} \ket{\psi_2 }$ \\ 
    $\ket{\psi_4 } = \prod_{e=1}^z \left(GW^{-2}\right)_{c,h_e} \prod_{j=1}^z W_{h_j} \ket{\psi_3 }$ \\
    $\ket{\psi_5 } = \prod_{j=z}^1 \mcx{d}_{m_{d  (j-1) + 1 } m_{d(j-1)+2}\ldots m_{d j}, h_j} \NOT_{h_j}  \ket{\psi_4 }$ \\ 
    $\ket{\psi_6 } = \h_c\prod_{j=n}^1  \XOR_{t_j, m_j}\NOT_{m_j} \ket{\psi_5 }$\\
    Measure qubit $\ket{c}$\\ 
    \If{$c == 0$}{
        Measure the memory register to obtain the desired state.
    }
    \caption{EP-PQM retrieval algorithm --- fault-tolerant implementation.}
    \label{alg:recover_ours}
\end{algorithm*}

\paragraph{NISQ implementation.}
Akin to the NISQ implementation~\cite{sousa2020parametric}, we decided to store the input pattern classically. This implementation requires only three registers: $m$,  $c$, and $h$, with $n$, $1$, and $z$ qubits, respectively:
\begin{equation}\label{eq:eppqm_retreival_registrers_hybrid}
    \underbrace{m_1 m_2 \ldots m_n}_{m}; \underbrace{c}_{c};\underbrace{h_1 h_2 \ldots h_z}_{h}.
\end{equation}
The initial quantum state for retrieval of information is
\begin{equation}\label{eq:hybrid-quantum-eppqm}
\begin{split}
\ket{\psi_0}_6 &= \frac{1}{\sqrt{\npat}} \sum_{k = 1}^{\npat} \ket{m_1^k m_2^k \ldots m_n^k;0;h_1 h_2 \ldots h_z}.
\end{split}
\end{equation}

Also, to further reduce the complexity of the circuit, we removed the $\NOT$ gates from Steps 3 and 5 of Algorithm~\ref{alg:recover_ours}. As in Algorithm~\ref{alg:recover_ppqm}, this results in measuring $\ket{m}$ when $\ket{c}$ is in state $\ket{1}$. As a result, Algorithm~\ref{alg:recover_ours} changes to Algorithm~\ref{alg:recover_ours_hybrid}. 
 
 \begin{algorithm}[t]
 \SetKwInOut{Input}{input}
    \SetKwInOut{Output}{output}
    \Input{The initial state $\ket{\psi_0}_6$ given in Eq.~\eqref{eq:hybrid-quantum-eppqm}} 
    \Output{1) The value of $c$ and 2) one of the stored patterns that has minimum $D$ with the target pattern in a given run}
     $\ket{\psi_1 } = \h_c\ket{\psi_0}_6$ \\

    \ForEach{$\tau_j \in T$}{ 
        
    \If{$\tau_j == 0$}{
    
        $\ket{\psi_2 } = \NOT_{m_j} \ket{\psi_1 }$ \\
        }
    \Else{
    $\ket{\psi_2 } =  \ket{\psi_1 }$
    }
    }
   
    $\ket{\psi_3 } = \prod_{j=1}^z \mcx{d}_{m_{d  (j-1) + 1 } m_{d(j-1)+2}\ldots m_{d j}, h_j} \ket{\psi_2 }$ \\ 
    
    $\ket{\psi_4 } = \prod_{e=1}^z \left(GW^{-2}\right)_{c,h_e} \prod_{j=1}^z W_{h_j} \ket{\psi_3 }$ \\
    
    $\ket{\psi_5 } = \prod_{j=z}^1 \mcx{d}_{m_{d  (j-1) + 1 } m_{d(j-1)+2}\ldots m_{d j}, h_j}  \ket{\psi_4 }$ \\ 
    
\ForEach{$\tau_j \in T$}{ 
        
    \If{$\tau_j == 0$}{
    
        $\ket{\psi_6 } = \NOT_{m_j} \ket{\psi_5 }$ \\
        }
    \Else{
    $\ket{\psi_6 } =  \ket{\psi_5 }$
    }
    }
    $\ket{\psi_7 } = \h_c\ket{\psi_6}$\\
    Measure qubit $\ket{c}$\\ 
    \If{$c == 1$}{
        Measure the memory register to obtain the desired state.
    }
    \caption{EP-PQM retrieval algorithm --- NISQ implementation.}
    \label{alg:recover_ours_hybrid}
\end{algorithm}

\subsection{Post-processing on classical computer}
The post-processing for EP-PQM is the same as that for PQM and P-PQM. We refer the reader to Section~\ref{sec:post-processing} for details.

\subsection{Implementation}\label{sec:implementation}

NISQ version of the EP-PQM can be implemented on any modern NISQ architecture. We introduce a reference implementation on QisKit~\cite{Qiskit}, a Python-based open-source software development kit for coding in OpenQASM and leveraging the IBM QCs. The code is given in~\cite{dat:github}; it is based on our PQM-based string comparison approach~\cite{khan2021string}.

Similar to the QisKit Aqua~\cite{AquaAlgo12:online} library, we wrap OpenQASM invocations into a Python class so that a programmer without any QC coding experience can leverage the algorithm from any Python program. The code can be executed in a simulator on a personal computer or on the actual IBM QC.

\subsection{Complexity analysis}
\subsubsection{Space complexity (qubits count)}\label{sec:space_complexity}
As discussed above, all the algorithms operate on $n$-bit patterns, but $n$ depends on the value of $a$ and the type of encoding (one-hot or label). Suppose we want to encode a pattern containing $z$ features with $a$ distinct attributes.

The case when $a=1$ is trivial because $D=0$ for any string.
For the case when $a>1$, formalizing Examples~\ref{ex:binary} and~\ref{ex:multi}, we may say that for PQM and P-PQM, which have to use one-hot encoding,
\begin{equation}\label{eq:n_ppqm}
n = n_o =
  \begin{cases}
    z       & \quad \text{if } a=2 \\
    za  & \quad \text{if } a > 2
  \end{cases},
\end{equation}
and for EP-PQM, which can use label encoding,
\begin{equation}\label{eq:n_eppqm}
n = n_l = z \attrspace  \quad \text{if } a \ge 2.
\end{equation}
In other words, when $a=2$, all algorithms require $n=z$ bits to represent a pattern, and when $a>2$, EP-PQM requires fewer bits and qubits. 

\paragraph{Fault-tolerant implementation.}
On an ideal QC, based on Eqs.~\eqref{eq:pqm_storage_register} and~\eqref{eq:pqm_retrieve_registry}, PQM and P-PQM require $2n + 2$ qubits for storage and $2n + 1$ qubits for retrieval, making a total of $4n + 3$ qubits~\cite{pqmct}. In practice, one can perform the storage and retrieval algorithms for PQM and P-PQM on a single circuit. This will lead to a reduction in the number of qubits to $2n + 2$~\cite{sousa2020parametric}.

In EP-PQM, we stick to the conceptually similar approach and, based on Eqs.~\eqref{eq:hybrid_storage_register} and~\eqref{eq:eppqm_retrieve_registry}, use $2n + z + 1$ qubits. In the fault-tolerant implementation, EP-PQM requires $2n+2$ qubits for storage and $2n+z+1$ for retrieval. During storage procedure, we need register $\ket{u}$ with two qubits. During retrieval, register $\ket{u}$ is not needed, thus its qubits can be reassigned. We can reuse one of these qubits for register $\ket{c}$ and one qubit as part of the register $\ket{h}$. Thus, we need only $2n+z+1$ qubits in total. 

Based on Eqs.~\eqref{eq:n_ppqm} and~\eqref{eq:n_eppqm}, in the $a=2$ case, PQM and  P-PQM will require $2z+2$ bits while EP-PQM will need $3z+1$ bits. When $a>2$ the savings start to emerge.
EP-PQM needs  $z\left(2\attrspace+1\right) + 1$ qubits, while PQM and P-PQM encoding need  $2za+2$ qubits.
Asymptotatically, this gives a reduction from  $O(za)$, in the PQM and P-PQM cases, to $O\left(z \log_2(a)\right)$, in the EP-PQM case.

\paragraph{NISQ implementation.} 
If we use one circuit for both the storage and retrieval algorithms, P-PQM's NISQ implementation\footnote{As mentioned in Section~\ref{sec:orig_retrieval}, PQM does not have a NISQ implementation.}, based on Eqs.~\eqref{eq:pqm_storage_register} and~\eqref{eq:ppqm_retreival_registrers}, requires $2n + 2$ qubits. EP-PQM, based on Eqs.~\eqref{eq:hybrid_storage_register} and~\eqref{eq:eppqm_retreival_registrers_hybrid}, uses $n + z + 1$ qubits (as in the fault-tolerant implementation above, we reassign qubits).

Based on Eqs.~\eqref{eq:n_ppqm} and~\eqref{eq:n_eppqm}, if $a=2$, EP-PQM will require one less qubit than P-PQM: $2z+1$ instead of $2z+2$.
And when $a>2$, P-PQM require $2za+2$ qubits, while EP-PQM needs only $z\left(\attrspace+1\right) + 1$. Asymptotatically, this again gives a reduction from $O(za)$ to $O\left(z \log_2(a)\right)$.

\paragraph{Summary.} In the fault-tolerant implementation, EP-PQM requires lesser number of bits and qubits than PQM and P-PQM when $a>2$. In the NISQ implementation, EP-PQM is more efficient than PQM and P-PQM for all values of $a$.

\subsubsection{Time complexity: overall number of gates}
Our algorithm's time complexity is proportional to the depth of the circuit for storage and retrieval. For now, we assume that different gates have the same time complexity (we will examine each gate type in Section~\ref{sec:specific_gates}).

\paragraph{Fault-tolerant implementation.}
In the fault-tolerant implementation, all three algorithms use Algorithm~\ref{alg:pqm_storing} to store the data, which requires $O(rn)$ gates.

PQM and P-PQM retrieval, based on Algorithm~\ref{alg:recover_orig}, need $O(n)$ gates. EP-PQM retrieval, based on Algorithm~\ref{alg:recover_ours}, needs  $O(n+z)$ gates. 

Consequently, the combined number of gates for storage and retrieval is $O(rn)$ for all three algorithms.
As in the case of space complexity, discussed above, the savings are driven by the encoding schema. Since for one-hot encoding $n=za$ and for label encoding $n=z \attrspace$, the number of gates for PQM and P-PQM is $O(rza)$, while for EP-PQM it is only $O\left(rz\log_2(a)\right)$. 

\paragraph{NISQ implementation.} 
PQM does not have a NISQ implementation, thus we focus on P-PQM and EP-PQM. The storage part of P-PQM, as in the fault-tolerant case, is  governed by Algorithm~\ref{alg:pqm_storing}, which, as discussed above, requires $O(rn)$ gates. EP-PQM storage is given in Algorithm~\ref{alg:ep-pqm_storing}; it still needs $O(rn)$ gates. 
As in the fault-tolerant case, the savings will come from the encoding schema, yielding $O(rza)$ gates for P-PQM and $O\left(rz\log_2(a)\right)$ gates for EP-PQM.

\paragraph{Summary.}  In both fault-tolerant and NISQ cases, EP-PQM becomes more efficient than PQM and P-PQM with the growth of $a$.

\subsubsection{Specific gates}\label{sec:specific_gates}
Complexity-wise, not all gates are created equal. Some gates increase complexity more than others. For example, on a NISQ device, $\mcx{2}$ gate requires more native gates than $\NOT$ gate, which increases complexity. 

\paragraph{Storage.} Algorithm~\ref{alg:pqm_storing} governs the data storage procedures for all three algorithms, except for the EP-PQM NISQ implementation, which is governed by Algorithm~\ref{alg:ep-pqm_storing}. Table~\ref{tbl:gate_count_storage} shows how many gates are required to implement these algorithms.  When comparing the number of individual gates, we have to be mindful of the encoding, as $n$ varies as per Eqs.~\eqref{eq:n_ppqm} and~\eqref{eq:n_eppqm}, where $n_o$ and $n_l$ denote $n$ in one-hot and label encoding cases, respectively.  

The parameter $\gamma \in [0,1]$ denotes a fraction of 1-bits in the database. 
The largest number of gates will be needed when $\gamma=1$.

Algorithm~\ref{alg:ep-pqm_storing} requires fewer control gates than Algorithm~\ref{alg:pqm_storing}. Table~\ref{tbl:gate_count_storage} shows that it does not need any $\mcx{2}$ gates. The number of $\XOR$ gates will vary with $n$, $r$, and $\gamma$. Let us analyze this variation by exploring under what conditions the number of $\XOR$ gates of EP-PQM is smaller than the number of P-PQM gates. That is, when does the following inequality hold:
\begin{equation}\label{eq:ineq_cnot_storage}
2 n_o r > 2 \gamma n_l r?    
\end{equation}

\begin{table}[ht]
    \centering
    \caption{Gate count for storage algorithms.}\label{tbl:gate_count_storage}
    \begin{tabular}{@{}lcc@{}}
    \toprule
    & All Fault-tolerant and P-PQM NISQ & EP-PQM NISQ  \\ 
    Gate & Algorithm~\ref{alg:pqm_storing} & Algorithm~\ref{alg:ep-pqm_storing} \\ \midrule
    $\XOR$       & $ 2n_or$    & $2 \gamma n_l r$ \\
    $\mcx{2}$    & $2n_or$     & $0$               \\
    $\mcx{n}$    & $2r$    & $2r$             \\
    $\NOT$       & $2n_or$            & $2 (1-\gamma) n_lr$ \\
    $\CS$      & $r$                & $r$\\
    \bottomrule
    \end{tabular}
\end{table}

For $a=2$ and using Eqs.~\eqref{eq:n_ppqm} and~\eqref{eq:n_eppqm}, the inequality~\eqref{eq:ineq_cnot_storage} becomes
$$ 2 n_o r > 2 \gamma n_l r \; \Rightarrow \;  2zr > 2 \gamma zr   \; \Rightarrow \;  1 > \gamma .$$
This inequality never holds, as $\gamma \le 1$. However, P-PQM and EP-PQM have the same number of $\XOR$ gates when $\gamma = 1$.

In the $a>2$ case,  Eq.~\eqref{eq:ineq_cnot_storage} becomes
\begin{equation*}
     2 n_o r  > 2\gamma n_l r  \; \Rightarrow \;  za > \gamma z \attrspace \; \Rightarrow \; a  > \gamma \attrspace  \; \Rightarrow \; \frac{a}{\attrspace} > \gamma. 
\end{equation*}
This inequality holds when $a>2$ for all $\gamma$. Thus, EP-PQM requires lesser number of $\XOR$ gates when $a>2$.

To understand $\NOT$ gates count, we need to analyze the following inequality:
\begin{equation}\label{eq:ineq_x_storage}
2 n_o r > 2 (1-\gamma) n_l r.   
\end{equation}

When $a=2$, using Eqs.~\eqref{eq:n_ppqm} and~\eqref{eq:n_eppqm}, Eq.~\eqref{eq:ineq_x_storage} simplifies to
$$2 n_o r > 2 (1-\gamma) n_l r  \; \Rightarrow \; 2zr > 2 (1-\gamma) zr  \; \Rightarrow \; 1 > 1 - \gamma.$$
This implies that P-PQM and EP-PQM have the same number of $\NOT$ gates when $\gamma = 0$. For other values of $\gamma$ EP-PQM outperforms P-PQM (when $a=2$).

Let us now explore $a>2$ case, where  Eq.~\eqref{eq:ineq_x_storage} becomes
\begin{equation}\label{eq:ineq_x_storage_details}
     2 n_o r > 2 (1-\gamma) n_l r  \; \Rightarrow \; za > (1-\gamma) z \attrspace  \;  \Rightarrow \; \frac{a}{\attrspace} + \gamma > 1. 
\end{equation}
This inequality will hold for all $\gamma$. Thus,  EP-PQM needs lesser number of $\NOT$ gates than P-PQM (when $a>2$).

In summary, we can say that EP-PQM will use the same number of gates as P-PQM when $a=2$ and $\gamma = 0$ (which is an extreme and rare case). For all other scenarios, EP-PQM will need lesser number of $\NOT$ gates than P-PQM.

\paragraph{Retrieval.} 
The comparison of the number of specific gate required to retrieve data based on Algorithms~\ref{alg:recover_orig}, \ref{alg:recover_ppqm}, \ref{alg:recover_ours}, and \ref{alg:recover_ours_hybrid} is given in Table~\ref{tbl:gate_count_retrieval}. The parameter $\delta \in [0,1]$ denotes the fraction of the bits in the input pattern that are 1's. 

\begin{table}[ht]
    \centering
    \caption{Gate count for retrieval algorithms. (P-)PQM stands for ``PQM and P-PQM''.}\label{tbl:gate_count_retrieval}
    \begin{tabular}{@{}lcccc@{}}
    \toprule
         & \multicolumn{2}{c}{Fault-tolerant} & \multicolumn{2}{c}{NISQ} \\ \cmidrule(l){2-3} \cmidrule(l){4-5}
         & (P-)PQM & EP-PQM & P-PQM & EP-PQM \\
    Gate & Algorithm~\ref{alg:recover_orig} & Algorithm~\ref{alg:recover_ours} &  Algorithm~\ref{alg:recover_ppqm} &  Algorithm~\ref{alg:recover_ours_hybrid} \\ \midrule
    $\XOR$       & $2n_o$  & $2n_l$ & $0$ & $0$\\
    $\mcx{d}$    &   $0$    & $2z$  & $0$ & $2z$\\
    $\NOT$       &  $2n_o$ & $2n_l + 2z$  & $2 \delta n_o$ & $2 (1-\delta) n_l$   \\
    $U$ or $W$   & $n_o$ & $z$  & $n_o$ & $z$ \\ 
    $GU$ or $GW$ & $n_o$ & $z$  & $n_o$ &  $z$ \\ 
    \h & $2$ & $2$  & $2$ & $2$ \\ 
    \bottomrule
    \end{tabular}
\end{table}

A few observations. Table~\ref{tbl:gate_count_retrieval} suggests that the retrieval phase requires lesser number of gates than storage phase (as it is independent of $r$). NISQ versions of the algorithms require the same or lesser number of gates than their fault-tolerant counterparts; this is expected as NISQ approaches are designed with the focus on performance.

For both fault-tolerant and NISQ, EP-PQM requires $2z$ additional $\mcx{d}$ gates than PQM and P-PQM. EP-PQM needs $a$ times less $W$ and $GW$ gates than corresponding PQM and P-PQM counterparts ($U$ and $GU$) for $a>2$.

For the \emph{fault-tolerant implementations}, when $a=2$, PQM and EP-PQM will need the same number of $\XOR$ gates (based on Eqs.~\ref{eq:n_ppqm} and~\ref{eq:n_eppqm}). When $a>2$, EP-PQM will need lesser number of $\XOR$ gates:
$$ 2 n_o > 2 n_l \; \Rightarrow \; 2za > 2z \attrspace \; \Rightarrow \; a > \attrspace .$$

For the $\NOT$ gates, when $a=2$, based on Eqs.~\eqref{eq:n_ppqm} and~\eqref{eq:n_eppqm}, P-PQM is more efficient than EP-PQM:
$$ 2n_o > 2n_l + 2z \; \Rightarrow \; 2z > 4z. $$
When $a>2$, the inequality becomes
\begin{equation*}
        2n_o > 2n_l + 2z \;  \Rightarrow \; 2za > 2z \attrspace + 2z  \;  \Rightarrow \;  a - \attrspace > 1.
\end{equation*}
This inequality holds for $a<4$. Thus, EP-PQM  needs lesser number of  $\NOT$ gates when $a \ge 4$.

For the \emph{NISQ implementations}, the number of $\NOT$ gates will vary with $\delta$. For the $a=2$ case
\begin{equation*}
   2 \delta n_o > 2 (1-\delta) n_l \; \Rightarrow \; 2 \delta z > 2 (1-\delta) z  \; \Rightarrow \;   \delta > (1-\delta). 
\end{equation*}
Thus, for $a=2$, if $\delta < 0.5$ then P-PQM needs lesser number of gates than EP-PQM, and if $\delta > 0.5$~--- vice versa. If $\delta = 0.5$, the number of gates is identical.

\begin{figure}[ht]
    \centering
    \includegraphics[width=\columnwidth]{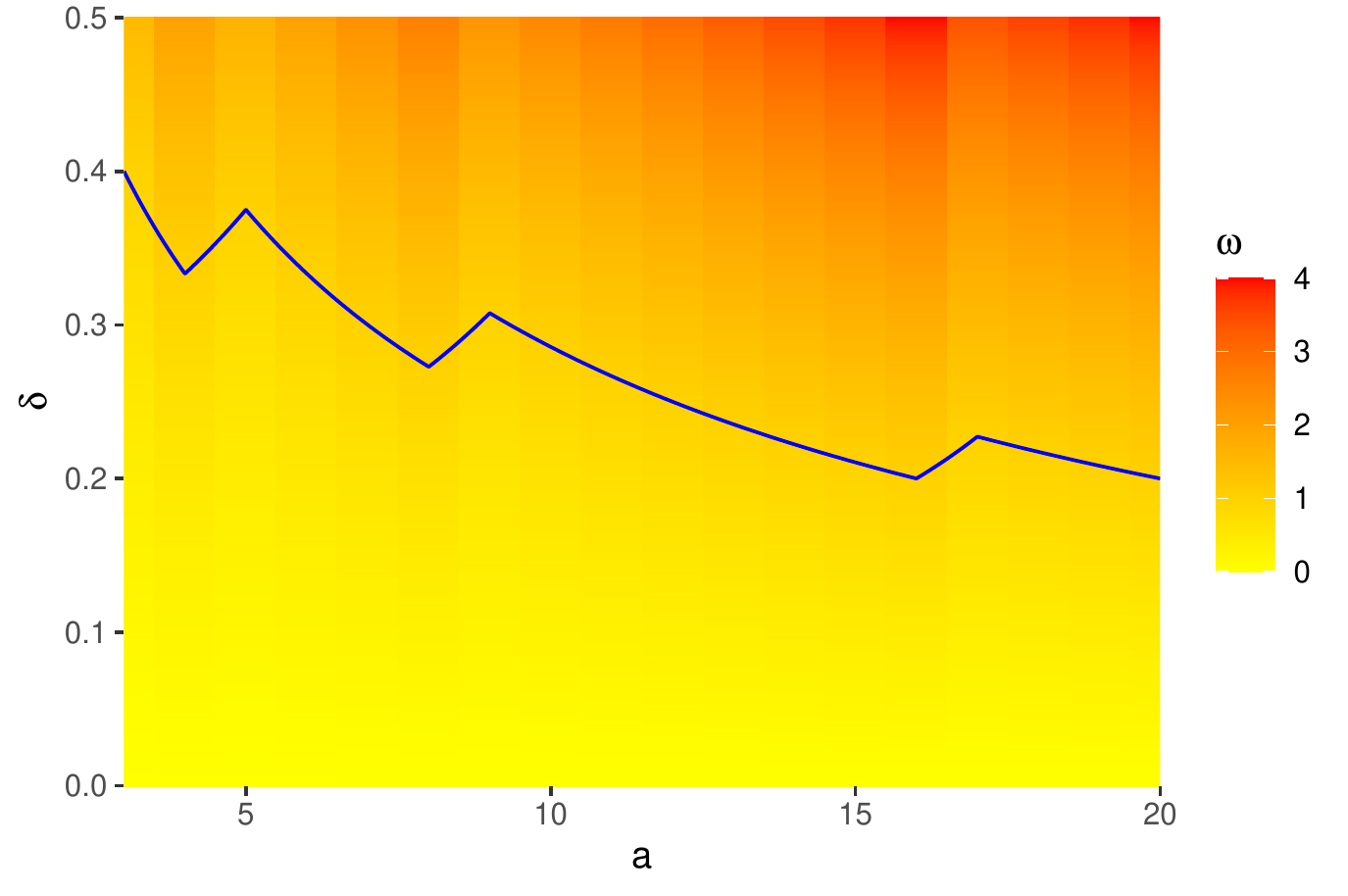}
    \caption{The z-axis of this contour plot depicts the values of $\omega$, defined in Eq.~\eqref{eq:ineq_x_retrieval_details_a_gt_2}. Blue line denotes the boundary where $\omega = 1$.  Above the blue line, P-PQM requires more $\NOT$ gates than EP-PQM; below it, P-PQM requires fewer $\NOT$ gates.
    }
    \label{fig:ineq_x_retreival}
\end{figure}

For the $a>2$ case, EP-PQM will need lesser number of $\NOT$ gates than P-PQM when
\begin{equation}\label{eq:ineq_x_retrieval_details_a_gt_2}
2 \delta n_o > 2 (1-\delta) n_l  \;  \Rightarrow \; \delta a > (1-\delta) \attrspace  \;  \Rightarrow \; \underbrace{\frac{ \delta}{1-\delta} \frac{a}{\attrspace}}_{\omega} > 1.
\end{equation}
The values of the inequality's left-hand-side are given in Figure~\ref{fig:ineq_x_retreival}. The inequality holds for different values of $\delta$ (in general, $\delta$ non-monotonically decreases with the increase of $a$). In other words, as $a$ increases, EP-PQM will require fewer $\NOT$ gates for a wider variety of input patterns. For example, when $a=3$ EP-PQM will require fewer $\NOT$ gates for any input pattern with more than 40\% of bits set to 1, and when $a=16$~--- for any input pattern with more than 20\% of bits set to 1.

\subsection{Remarks}
\subsubsection{Label encoding}
The principles of label encoding on a classical or a quantum computer are the same. Label encoding can, for instance, be applied directly to nominal and ordinal data. This method can also be applied to numeric data if they are mapped to labelled intervals.

\subsubsection{Attributes}
So far, we were operating under the implicit assumption that each feature in an input dataset has the same number of attributes. However, in practice, this is often not the case. Our current (implicit) workaround is to set the number of characters in the alphabet to the maximum distinct number of attributes of any feature in the dataset.

For example, one feature may have two distinct attributes and another feature~--- 100 distinct attributes. In this case we will use seven bits (i.e., $\ceil{\log_2(100)}$)  to represent each of the features.

To use only the minimal number of bits needed to represent each feature, we need to alter
Steps~5 and~7 of Algorithms~\ref{alg:ep-pqm_storing}. Specifically, we need to replace\footnote{Gate replacement is all that is needed for ML classification task. To use this approach for computing the value of $D$ on a classical computer, we also need to alter the trigonometric formulas~\cite[Eqs. 2 and 3]{khan2021string}. } the $\mcx{d}$ gate for $j$-th feature with $\mcx{d\textsubscript{j}}$, where $d_j$ is the number of bits needed to represent an attribute of the $j$-th feature.

\subsubsection{Other use cases}
In this paper, we focus on the Quantum ML classification use case. However, the same principles can be applied to any use case requiring encoding of data with multiple attributes. 

For example, by using this approach, we can efficiently codify strings where characters are drawn from an alphabet with a large number of characters. A string may represent anything from mRNA sequences in bioinformatics to software log records in software engineering~\cite{khan2021string}.

\section{Experiments}\label{sec:exp}
In order to compare P-PQM with EP-PQM, we show the number of qubits and gates necessary to perform classification on five datasets from the UCI Machine Learning Repository~\cite{dua2017uci} (which were also used by the authors of P-PQM~\cite{sousa2020parametric}). Summary statistics for the datasets are given in Table~\ref{tbl:datasets}. Data are encoded using one-hot encoding in the P-PQM case, whereas label encoding is used in the EP-PQM case. 
We do not measure the accuracy of the models in our experiments because PQM and EP-PQM yield the same accuracy (since $D$ is the same in both cases).

\begin{table}[ht]
\caption{Datasets' description (based on 90\% of the observations) of the class/label having the maximum number of observations.}\label{tbl:datasets}
\begin{center}
\begin{tabular}{@{}llrrr@{}}
\toprule
Dataset       & Class/Label name    & $r$ & $z$ & $a$ \\ \midrule
Balance Scale~\cite{uci:balancescale, paper:balancescale} & R        & 262                 & 4               & 5                \\
Breast Cancer~\cite{uci:breastcancer, paper:breastcancer} & 2        & 412                 & 9               & 11               \\
SPECT Heart~\cite{uci:spectheart, paper:spect}   & 1        & 36                  & 22              & 2                \\
Tic-Tac-Toe Endgame~\cite{uci:tictactoe, paper:tictactoepaper}      & positive & 558                 & 9               & 3                \\
Zoo~\cite{uci:zoo, paper:zoopaper}            & 1        & 40                  & 16              & 6               \\ \bottomrule
\end{tabular}
\end{center}
\end{table}

Sousa et al.~\cite{sousa2020parametric} tested the performance of ML models using a 10-fold cross-validation approach, which assesses how well the training results generalize to previously unexplored data. In this approach, a dataset is divided into ten equal subsets. Then, nine subsets are used to train the model and one to validate the training results. The process is repeated ten times, with each of the subsets being used for validation only once.

In order to mimic this approach, we sample 90\% of observations (i.e., patterns) from each dataset, referred to as $b$. The observations in the dataset belong to different classes.  Let us assume the dataset has $l$ distinct classes/labels. We group $b$ observations by $l$ labels, creating $l$ databases of observations. We then focus on of the $l$ databases with the most observations (as it represents the hardest task), since it will require the deepest quantum circuit for a given dataset. Let us look at a toy example.

\begin{exmp}
Consider a dataset with three classes/labels and 1000 observations. We sample 90\% of observations, i.e., $b=900$. Suppose 500 observations belong to label 1, 300 observations~--- to label 2, and 100 observations~--- to label 3. We will focus only on the database of observations for label 1, as it is the largest.
\end{exmp}

We implement the quantum circuit needed to store and retrieve data using P-PQM (Algorithms~\ref{alg:pqm_storing} and~\ref{alg:recover_ppqm}) and EP-PQM (Algorithms~\ref{alg:ep-pqm_storing} and~\ref{alg:recover_ours_hybrid}). All quantum algorithms are implemented using Qiskit v0.33.1~\cite{Qiskit}. The results for the quantum-circuit-based implementation are shown below; the results for the Quantum Simulator backend can be found in Appendix~\ref{sec:qsim}.

For each dataset's database with the most observations, we compute the quantum circuit depth and the number of gates needed to execute NISQ storage and retrieval algorithms. Finally, we perform a shallow decomposition of this circuit (by calling ``decompose()'' methods of the QisKit quantum circuit) and obtain the set of gates (shown in Table~\ref{tbl:low_gates_decomposed}).

Table~\ref{tbl:datasets} contains high-level statistics for the datasets under study. The datasets have a diverse structure. For instance, SPECT Heart dataset has only two features and will be stored using binary variables for both P-PQM and EP-PQM. The rest of the algorithms have $a$ between $3$ and $11$. Thus, for P-PQM we will encode the attributes using one-hot encoding and for EP-PQM~--- using label encoding.

Table~\ref{tbl:space_depth} shows the number of qubits and quantum circuit depths. In addition, we compute relative resource savings while treating the resources needed for P-PQM as a baseline. The number of qubits used by P-PQM and EP-PQM in SPECT Heart is the same\footnote{For this implementation of P-PQM and EP-PQM, to simplify the code, one qubit from $\ket{u}$ was reassigned to $\ket{c}$, but the second qubit of $\ket{u}$ was not reassigned to $\ket{h}$. EP-PQM therefore requires one extra qubit in comparison with the theoretical requirement: $2n+2$ instead of $2n+1$.}, which is expected as $a=2$.
Label encoding reduces the number of qubits required by 48\% to 77\% in the rest of the datasets. 

Peculiarly, quantum circuit depth is reduced for all datasets by 60\% to 96\%, showing the effectiveness of EP-PQM even for datasets with $a=2$. The savings are even more evident after decomposition: the decomposed quantum circuit depth is reduced by 94\% to 99\%. These additional savings can be explained by the fact that many gate types (especially multi-controlled ones) require a lot of native gates to be implemented, thus increasing the quantum circuit depth.

\begin{table*}[ht]
\caption{Comparison of the number of qubits and quantum circuit depth for class/label with maximum number of observations. A comma separates groups of thousands.}\label{tbl:space_depth}
\begin{center}
\resizebox{\linewidth}{!}{
\begin{tabular}{@{}lrrrrrrrrr@{}}
\toprule
                              & \multicolumn{3}{c}{Number of Qubits} & \multicolumn{3}{c}{Quantum Circuit Depth} & \multicolumn{3}{c}{Decomposed Quantum Circuit Depth} \\  \cmidrule(r){2-4} \cmidrule(r){5-7} \cmidrule(r){8-10}
Dataset                       & P-PQM    & EP-PQM    & Savings    & P-PQM     & EP-PQM    & Savings & P-PQM     & EP-PQM    & Savings   \\  \midrule
Balance Scale & 42       & 18        & 57\%      & 12,338     & 2,899      & 77\%   &  108,510  &	 4,515 & 96\%   \\
Breast Cancer & 200      & 47        & 77\%      & 84,563     & 9,776     & 88\% &  821,728 &	 12,407 &	98\%     \\
SPECT Heart      & 46       & 46        & 0\%       & 1,862      & 747       & 60\%  & 16,390 &	 982 &	94\%    \\
Tic-Tac-Toe Endgame      & 56       & 29        & 48\%      & 34,069     & 8,478     & 75\%  & 309,188 &	 11,847 &	96\%    \\
Zoo            & 194      & 66        & 66\%      & 8,060      & 334       & 96\%  & 77,554 &	 629 &	99\%    \\ \bottomrule
\end{tabular}
}
\end{center}
\end{table*}

As discussed in Section~\ref{sec:specific_gates}, to understand efficiency of the algorithms, we need to asses the count of specific gates. This information is given in Table~\ref{tbl:low_gates}. 

We will use \texttt{typewriter font} to reference the gates used in the QisKit's implementation of the circuit. A number of gates are readily available. The $\XOR$ gate is implemented using \qk{cx}.
The $\mcx{\textalpha}$ gates, where $\alpha$ is the number of control qubits, will be implemented using \qk{cx} when there is one control qubit, \qk{ccx} when there are  two control qubits,  \qk{mcx} when there are 3 or 4 control qubits and  \qk{mcx\_gray} when there are more than or equal to 5 control qubits. The $\h$ gate is implemented using \qk{h}, and $\NOT$ gate is implemented using \qk{x}. 

According to the table, EP-PQM significantly reduces the number of control gates \qk{cx}, \qk{ccx}, and \qk{mcx\_gray} by 58.7\% to 100.0\%. For three datasets in EP-PQM with $a \ge 4$,  a small number of \qk{mcx} control gates ($2z$, from 8 to 32) were added to implement $\mcx{d}$ for three datasets with $a \ge 4$; for P-PQM, no \qk{mcx} gates were required. 

The EP-PQM approach results in significant reductions in other gates (namely, \qk{unitary} and \qk{cunitary}\footnote{  \qk{unitary} denotes a custom unitary gate and \qk{cunitary} is a controlled version of a custom unitary gate.} QisKit simulator gates which cover algorithmic $U$, $GU$, $W$, $CW$, and $\CS$ gates) from 3.1\% to 90.9\% for all datasets except for SPECT Heart dataset where the number of gates was identical to the P-PQM algorithm.

For all datasets, EP-PQM reduces the number of $\NOT$ gates by 53.2\% to 76.2\%.

The count of gates obtained using shallow decomposition of the quantum circuit is given in Table~\ref{tbl:low_gates_decomposed}. Decomposition of the circuit of the Breast Cancer dataset yielded gates \qk{c3sx}, \qk{cu1}, \qk{rcccx}, and \qk{rcccx\_dg} not found in circuits of other datasets. To facilitate comparison with other datasets, we further decomposed these gates using ``decompose([`c3sx', `rcccx', `rcccx\_dg']).decompose(`cu1')''. This results in all datasets containing the same gates.

The savings are even more pronounced. For $\qk{cu}$, $\qk{cx}$, $\qk{h}$, $\qk{measure}$, $\qk{t}$, $\qk{tdg}$ and $\qk{u3}$ the number of gates reduced by 32.1\% to 100.0\%. Also, the number of phase gates, \qk{p}, reduced by 66.7\% and 90.9\% for Tic-Tac-Toe Endgame dataset and Breast Cancer dataset respectively. 

The number of gates for \qk{p} in the Balance Scale dataset and Zoo dataset increased by 520.0\% and 416.7\% respectively and the number of \qk{u2} gates (where  \qk{u2} is a single qubit gate) increased by 7200.0\% for the Breast Cancer dataset. While the relative increase is significant, the absolute increase is relatively small~--- there are only a few hundred gates added.  The number of single qubit gates \qk{u1} increased from 0 to 774 (as above, this number is small). 

Overall, our experiments confirm our theoretical analysis and show that EP-PQM leads to a significant reduction in gate count compared to P-PQM. Consequently, EP-PQM is more applicable for NISQ devices due to a reduction in quantum circuit depth. 
\afterpage{
\clearpage
\thispagestyle{empty}
\begin{landscape}
\begin{table}[ht]
\caption{Gate count in the quantum circuit. P stands for P-PQM, EP --- for EP-PQM, and S --- for relative savings computed as $(\textrm{P} - \textrm{EP})/ \textrm{P}$. The gates used in QisKit's implementation of the quantum circuit are as follows: \qk{ccx} is a Toffoli gate $\mcx{2}$; \qk{cunitary} and \qk{unitary} are custom unitary gates (which our $U$, $GU$, $W$, $CW$, and $\CS$ gates translate to); \qk{cx} is a $\XOR$ gate; \qk{h} is a $\h$ gate; \qk{mcx} is a $\mcx{\textalpha}$ gate where $\alpha$ is 3 or 4; \qk{mcx\_gray} is a $\mcx{\textalpha}$ gate where $\alpha$ $\geq$ 5; \qk{measure} is the measurement of the qubit into classical bit; \qk{x} is a $\NOT$ gate; see QisKit documentation for details~\cite{Qiskit}. We deliberately leave QisKit gate names as-is to differentiate ``algorithmic gates'' from  ``implementation gates''.}\label{tbl:low_gates}
\resizebox{\linewidth}{!}{
\begin{tabular}{@{}lrrrrrrrrrrrrrrrrrrrrrrrrrrr@{}}
\toprule
                              & \multicolumn{3}{c}{\qk{ccx}}  & \multicolumn{3}{c}{\qk{cunitary}} & \multicolumn{3}{c}{\qk{cx}}   & \multicolumn{3}{c}{\qk{h}}    & \multicolumn{3}{c}{\qk{mcx}}  & \multicolumn{3}{c}{\qk{mcx\_gray}} & \multicolumn{3}{c}{\qk{measure}} & \multicolumn{3}{c}{\qk{unitary}} & \multicolumn{3}{c}{\qk{x}}    \\
                              \cmidrule(r){2-4} \cmidrule(r){5-7}  \cmidrule(r){8-10}  \cmidrule(r){11-13}  \cmidrule(r){14-16} \cmidrule(r){17-19} \cmidrule(r){20-22} \cmidrule(r){23-25} \cmidrule(r){26-28}
Dataset                       & P & EP & S & P   & EP  & S  & P & EP & S & P & EP & S & P & EP & S & P   & EP   & S  & P  & EP  & S  & P  & EP  & S  & P & EP & S \\ \midrule
Balance Scale & 10,480 &  0     & 100.0\% & 20      & 4       & 80.0\%   & 10,480 & 2104   & 79.9\%  & 2     & 2      & 0.0\%   & 0       & 8      & --         & 524     & 524      & 0.0\%    & 21     & 13      & 38.1\%   & 282    & 266     & 5.7\%    & 12,585 & 4,207   & 66.6\%  \\
Breast Cancer & 81,576 &  0     & 100.0\% & 99      & 9       & 90.9\%   & 81,576 & 8526   & 89.5\%  & 2     & 2      & 0.0\%   & 0       & 18     & --         & 824     & 824      & 0.0\%    & 100    & 37      & 63.0\%   & 511    & 421     & 17.6\%   & 89,011 & 21,200  & 76.2\%  \\
SPECT Heart      & 1,584  &  0     & 100.0\% & 22      & 22      & 0.0\%    & 1,584  & 654    & 58.7\%  & 2     & 2      & 0.0\%   & 0       & 0      & --         & 72      & 72       & 0.0\%    & 23     & 23      & 0.0\%    & 58     & 58      & 0.0\%    & 2,205  & 1,031   & 53.2\%  \\
Tic-Tac-Toe Endgame      & 30,132 & 18     & 99.9\%  & 27      & 9       & 66.7\%   & 30,132 & 6790   & 77.5\%  & 2     & 2      & 0.0\%   & 0       & 0      & --         & 1,116    & 1,116     & 0.0\%    & 28     & 19      & 32.1\%   & 585    & 567     & 3.1\%    & 40,195 & 13,328  & 66.8\%  \\
Zoo            & 7,680  &  0     & 100.0\% & 96      & 16      & 83.3\%   & 7,680  & 188    & 97.6\%  & 2     & 2      & 0.0\%   & 0       & 32     & --         & 80      & 80       & 0.0\%    & 97     & 49      & 49.5\%   & 136    & 56      & 58.8\%   & 8,993  & 3,757   & 58.2\% \\ \bottomrule
\end{tabular}
}
\end{table}

\begin{table}[ht]
\caption{Gate count in the decomposed quantum circuit. P stands for P-PQM, EP --- for EP-PQM, and S --- for relative savings. }\label{tbl:low_gates_decomposed}
\resizebox{\linewidth}{!}{
\begin{tabular}{lrrrrrrrrrrrrrrrrrrrrrrrrrrrrrrrrr}
\toprule
                    &  \multicolumn{3}{c}{\qk{cu}}                                                 &  \multicolumn{3}{c}{\qk{cx}}                                                 & \multicolumn{3}{c}{\qk{h}}                                                  & \multicolumn{3}{c}{\qk{mcu1}}                                               & \multicolumn{3}{c}{\qk{measure}}                                            & \multicolumn{3}{c}{\qk{p}}                                                  &  \multicolumn{3}{c}{\qk{t}}                                                  & \multicolumn{3}{c}{\qk{tdg}}                                                & 
                    \multicolumn{3}{c}{\qk{u1}}                                                 &
                    \multicolumn{3}{c}{\qk{u2}}                                                 & \multicolumn{3}{c}{\qk{u3}}                                                 \\
                    \cmidrule(r){2-4} \cmidrule(r){5-7}  \cmidrule(r){8-10}  \cmidrule(r){11-13}  \cmidrule(r){14-16} \cmidrule(r){17-19} \cmidrule(r){20-22} \cmidrule(r){23-25} \cmidrule(r){26-28} \cmidrule(r){29-31} \cmidrule(r){32-34} 
Dataset             &  \multicolumn{1}{c}{P} & \multicolumn{1}{c}{EP} & \multicolumn{1}{c}{S} & \multicolumn{1}{c}{P} & \multicolumn{1}{c}{EP} & \multicolumn{1}{c}{S} & \multicolumn{1}{c}{P} & \multicolumn{1}{c}{EP} & \multicolumn{1}{c}{S} & \multicolumn{1}{c}{P} & \multicolumn{1}{c}{EP} & \multicolumn{1}{c}{S} & \multicolumn{1}{c}{P} & \multicolumn{1}{c}{EP} & \multicolumn{1}{c}{S} & \multicolumn{1}{c}{P} & \multicolumn{1}{c}{EP} & \multicolumn{1}{c}{S} & \multicolumn{1}{c}{P} & \multicolumn{1}{c}{EP} & \multicolumn{1}{c}{S} & \multicolumn{1}{c}{P} & \multicolumn{1}{c}{EP} & \multicolumn{1}{c}{S} & \multicolumn{1}{c}{P} & \multicolumn{1}{c}{EP} & \multicolumn{1}{c}{S} & \multicolumn{1}{c}{P} & \multicolumn{1}{c}{EP} & \multicolumn{1}{c}{S} & \multicolumn{1}{c}{P} & \multicolumn{1}{c}{EP} & \multicolumn{1}{c}{S} \\ \midrule
Balance Scale       & 20                    & 4                      & 80.0\%              & 73,883                 & 2,739                   & 96.3\%                & 22,008                 & 1,064                   & 95.2\%                & 524                   & 524                    & 0.0\%                 & 21                    & 13                     & 38.1\%                & 20                    & 124                    & -520.0\%       & 41,920                 &                        & 100.0\%               & 31,440                 &                        & 100.0\%                 &                       &                        & --                                  & 2                     & 2                      & 0.0\%                 & 14,175                 & 5,781                   & 59.2\%                \\
Breast Cancer      & 99                    & 9                      & 90.9\%              & 571,855                 & 9,997                   & 98.3\%                & 164,800                & 1,972                   & 98.8\%                & 824                   & 824                    & 0.0\%                 & 100                   & 37                     & 63.0\%                & 99                    & 9                      & 90.9\%               & 326,304                &                        & 100.0\%               & 244,728                 &                        & 100.0\%                &                       &  774                      & --                                   & 2                     & 146                      & -7200.0\%                 & 91,580                 & 23,679                  & 74.1\%                \\
SPECT Heart         & 22                    & 22                     & 0.0\%            & 11,159                 & 725                    & 93.5\%                & 3,312                  & 144                    & 95.7\%                & 72                    & 72                     & 0.0\%                 & 23                    & 23                     & 0.0\%                 & 22                    & 22                     & 0.0\%       & 6,336                  &                        & 100.0\%               & 4,752                  &                        & 100.0\%                 &                       &                        & --                                  & 2                     & 2                      & 0.0\%                 & 2,441                  & 1,267                   & 48.1\%                \\
Tic-Tac-Toe Endgame & 27                    & 9                      & 66.7\%            & 21,2039                 & 8,013                   & 96.2\%                & 62,496                 & 2,268                   & 96.4\%                & 1,116                  & 1,116                   & 0.0\%                 & 28                    & 19                     & 32.1\%                & 27                    & 9                      & 66.7\%                  & 120,528                & 72                     & 99.9\%                & 90,396                 & 54                     & 99.9\%                  &                       &                        & --                                  & 2                     & 2                      & 0.0\%                 & 43,568                 & 16,683                  & 61.7\%                \\
Zoo              & 96                    & 16                     & 83.3\%              & 53,839                 & 715                    & 98.7\%                & 15,520                 & 224                    & 98.6\%                & 80                    & 80                     & 0.0\%                 & 97                    & 49                     & 49.5\%                & 96                    & 496                    & -416.7\%              & 30,720                 &                        & 100.0\%               & 23,040                 &                        & 100.0\%                    &                       &                        & --                               & 2                     & 2                      & 0.0\%                 & 9,327                  & 4,011                   & 57.0\%               \\ \bottomrule
\end{tabular}
}
\end{table}
\end{landscape}
\clearpage
}

\section{Summary}\label{sec:summary}
In this paper, we extend the PQM-based ML classification algorithm designed for use in NISQ devices. The original approach provided correct classification using one-hot encoding. We extend this approach to enable label encoding, which reduces space complexity (i.e., qubit count) from $O(za)$ to $O\left(z \log_2(a)\right)$ and decreases the number of gates in the quantum circuit from $O\left(rza\right)$ to $O\left(rz\log_2(a)\right)$. 

By simulating ML classification on five datasets (using QisKit QC Simulator) and analyzing the resulting circuit, we verified our theoretical analysis. Depending on the dataset, EP-PQM quantum circuit depth saving range between 60\% and 96\%. Similarly, EP-PQM reduces the corresponding decomposed quantum circuit depth between 94\% and 99\%. Qubit count was reduced by 48\% to 77\% for datasets with $a>2$.

A reduction in space requirements makes it possible to load larger datasets into a QC. Furthermore, reducing the number of gates helps speed up classification and decrease noise associated with deep quantum circuits.

\section*{Acknowledgements}
We are grateful to Natural Sciences and Engineering Research Council of Canada (NSERC) for financial support and Compute Canada for providing access to computers with large amounts of memory. We would like to acknowledge Canada’s National Design Network (CNDN) for facilitating this research, specifically through their member access to the IBM Quantum Hub at Institut quantique.
\appendix

\section{One Hot Encoding and $D$: Relationship}\label{sec:one-hot-and-d}
Suppose there are two patterns $X = x_1, x_2, \ldots, x_z$ and $Y = y_1, y_2, \ldots, y_z$, containing $z$ features with $a$ attributes. Further suppose that $X$ and $Y$ represented using one-hot encoding. Thus, $i$-th feature will be represented by $a$-bit string. 

Let the Hamming distance be denoted by $D$ and the Hamming distance computed when comparing patterns encoded using one-hot encoding be denoted by $D_{\textrm{one-hot}}$.
When performing pair-wise comparison, $D(x_i, y_i) = D_{\textrm{one-hot}}(x_i, y_i) = 0$, when $x_i$ and $y_i$ are identical. The similarity ends when $x_i \ne y_i$. By definition
\begin{equation}\label{eq:d}
    D(x_i, y_i) = 1,~\textrm{when}~x_i \ne y_i.
\end{equation}
However, by construction of the one-hot encoding, 
\begin{equation}\label{eq:d_one_hot}
    D_{\textrm{one-hot}}(x_i, y_i) = 2,~\textrm{when}~x_i \ne y_i.
\end{equation}
 That is, $x_i$ and $y_i$ will have a single bit set to $1$ at different positions, which requires two operations to convert $x_i$ into $y_i$.

Extrapolating Eqs.~\eqref{eq:d} and~\eqref{eq:d_one_hot} to $X$ and $Y$, given that $k$ features are different, 
$D(X, Y) = \sum_{i=1}^k{D(x_i, y_i)} = k$, while $D_{\textrm{one-hot}}(X, Y) = \sum_{i=1}^k{D_{\textrm{one-hot}}(x_i, y_i)} = 2k$. This implies that $$D(X, Y) = \frac{1}{2} D_{\textrm{one-hot}}(X, Y).$$

\section{Quantum Computing: Fundamentals}\label{sec:qc}
Here, we review the fundamentals of quantum computing needed to implement the algorithms described in this paper. See~\cite{nielsen_chuang_2010} for more information about quantum computing.

In quantum computing, a qubit represents the basic unit of information. It is formed by two-quantum states: one constitute the state $\ket{0} = \begin{bmatrix} 1 & 0 \end{bmatrix}^\intercal$ and the other the state $\ket{1} = \begin{bmatrix} 0 & 1 \end{bmatrix}^\intercal$. The vectors $\ket{0}$ and $\ket{1}$ which are the eigenvectors of the Pauli-$Z$ matrix are known as the computational basis. The others well known basis are the Pauli-$X$ and Pauli-$Y$ eigenvectors. Differently from a classical bit, a qubit can exist in a superposition state 
$\ket{\psi} =  \alpha \ket{0} + \beta \ket{1} = \begin{bmatrix} \alpha & \beta \end{bmatrix}^\intercal$,
where $\alpha$ and $\beta$ are, respectively, the amplitude of the qubit in state $\ket{0}$ and $\ket{1}$. The amplitudes are complex numbers which satisfy the normalization condition $\abs{\alpha}^2 + \abs{\beta}^2 = 1$. The state $\ket{\psi}$ is said to be in a superposition of states $\ket{0}$ and $\ket{1}$, which reflects the fact that the qubit can be in more than one basis state at a particular time with a given probability.

To process quantum information, a sequence of quantum gates and measurements must be performed on the qubits. The quantum circuit is formed by these operations. Quantum gates are unitary operations acting on single or multiple qubits. Quantum gates are responsible for evolving the state of the QC and forming the quantum algorithm solution. To learn the QC state, qubits are measured. The measurement operation transforms the qubit state $\ket{\psi}$ in a classical bit $0$ or $1$, which are measured with probability $|\alpha|^2$ or $|\beta|^2$, respectively. The quantum gates used in this paper are defined below.

The Hadamard gate, given by 
$$\h = \frac{1}{\sqrt{2}} \begin{bmatrix} 1 & 1\\ 1 & -1\\  \end{bmatrix}, $$
transforms the state $\ket{0} \to \frac{1}{\sqrt{2}}  (\ket{0} + \ket{1})$ and $\ket{1} \to \frac{1}{\sqrt{2}}(\ket{0} - \ket{1})$, which are both superpositions. If the state is already in a superposition, applying the Hadamard gate on it will revert it to $\ket{0}$ or $\ket{1}$, respectively.

Pauli-X gate is defined as
$$\NOT = \begin{bmatrix} 0 & 1\\ 1 & 0\\  \end{bmatrix}.$$
It is the $\NOT$ gate which flips the qubit state, transforming the state $\ket{0}$ to $\ket{1}$ and the state $\ket{1}$ to $\ket{0}$. 

Controlled-NOT (CNOT) gate, given by
$$\XOR = \begin{bmatrix} 1 & 0 & 0&0\\ 0 & 1 & 0&0\\ 0 & 0 & 0&1 \\ 0 & 0 & 1&0\end{bmatrix},$$
is a two-qubit gate, operating on control and target qubits. It applies the $\NOT$ gate to a target qubit whenever its control qubit is in state $\ket{1}$. In this context, we can interpret the $\XOR$ as the classic $\textsc{xor}$ gate. $\XOR$ gate can be extended to having $\alpha$ control qubits (\mcx{\textalpha}). In this case, the $\NOT$ gate is applied to the target qubit whenever each of the $\alpha$ control qubits are in state $\ket{1}$. The $\mcx{\textalpha}$ is $2^{\alpha+1} \times 2^{\alpha+1}$ unitary operator. All control gates have subscripts indicating the qubits on which they are applied, with the control qubits listed first. For example, in control gate $\mcx{\textalpha}_{m_1\cdots m_n,u_1}$ qubits $m_1\cdots m_n$ act as controls and qubit $u_1$
is a target.

Finally, the last operation in a quantum circuit is the qubit-state measurement. Indeed, when a quantum system is not measured, a qubit can be in a superposition of states of $\ket{0}$ and $\ket{1}$. However, after measurement, the qubit state collapses into either the state $\ket{0}$ or $\ket{1}$ with a probability of the absolute value of the amplitude squared.

\section{Circuit for Quantum Simulator}\label{sec:qsim}
We have also tried to transpile the circuit for the QisKit Simulator backend (namely, ``QasmSimulator''). To minimize computation efforts, no optimization of the circuit is performed (``optimization\_level = 0''). The transpilation, however, requires a substantial amount of memory: empirically, we found that $1$~TB of memory was insufficient for transpiling the circuits for our datasets. This was not expected, as the generation of data in Tables~\ref{tbl:space_depth}, \ref{tbl:low_gates}, and~\ref{tbl:low_gates_decomposed} required $\approx 3$~GB of memory. 

Therefore, to get a feel for the QasmSimulator-based circuits, we reduced the number of observation to two in each dataset, i.e., $r=2$. In spite of this simplification, we still ran out of memory for the Breast-Cancer-, Tic-Tac-Toe-, and Zoo-based circuits. However, we could produce the circuits for the Balance Scale and SPECT Heart datasets. 

Table~\ref{tbl:sim_gates_cnt} shows results of transpilation without optimization, and Table~\ref{tbl:sim_decomposed_gates_cnt} shows stats for decomposed versions of these circuits. According to the tables, the results are mixed. 

For Balance Scale dataset, the quantum circuit depth is three orders of magnitude smaller, resulting in overall savings of 99.6\%. In the original circuit, the biggest savings are associated with gates $\qk{cx}$ and $\qk{u1}$, while in the decomposed case, the biggest savings are associated with $\qk{cx}$ and $\qk{u3}$. Savings would be even greater with the increase in $r$.

For SPECT Heart dataset, the quantum circuit depth is reduced, but the reduction is marginal: 0.001\%, because the number of  $\qk{cx}$ and $\qk{u1}$ gates are almost identical for both approaches. This differs from the results that we saw in Tables~\ref{tbl:space_depth}, \ref{tbl:low_gates}, and~\ref{tbl:low_gates_decomposed}: there the quantum circuit depth was reduced by 60\% and 94\% for quantum circuit and decomposed quantum circuit, respectively. 

We conjecture that this stark difference may be due to inefficiencies in the Quantum Simulator's optimizer. Significant memory requirements indirectly support this conjecture.

\afterpage{%
\clearpage
\thispagestyle{empty}
\begin{landscape}

\begin{table}[ht]
\caption{Gate count in the quantum circuit for QasmSimulator when $r=2$.  P stands for P-PQM, EP --- for EP-PQM, and S --- for relative savings.}\label{tbl:sim_gates_cnt}
\resizebox{\linewidth}{!}{
\begin{tabular}{lrrrrrrrrrrrrrrrrrrrrr}
\toprule
              & \multicolumn{3}{c}{Quantum Circuit Depth}                                                     & \multicolumn{3}{c}{\qk{cx}}                                                                & \multicolumn{3}{c}{\qk{measure}}                                                  & \multicolumn{3}{c}{\qk{u1}}                                                                & \multicolumn{3}{c}{\qk{u2}}                                                        & \multicolumn{3}{c}{\qk{u3}}                                                        & \multicolumn{3}{c}{\qk{unitary}}                                                 \\
              \cmidrule(r){2-4} \cmidrule(r){5-7}  \cmidrule(r){8-10}  \cmidrule(r){11-13}  \cmidrule(r){14-16} \cmidrule(r){17-19} \cmidrule(r){20-22}
              & P                            & EP                         & S                          & P                            & EP                         & S                          & P                      & EP                      & S                          & P                            & EP                         & S                          & P                       & EP                      & S                          & P                       & EP                      & S                          & P                      & EP                     & S                          \\ \midrule
Balance Scale & 20,972,436 & 81,981 & 99.6\% & 12,583,496 & 49,274 & 99.6\% & 21 & 13 & 38.1\% & 12,583,540 & 49,276 & 99.6\% & 170 & 26 & 84.7\% & 145 & 61 & 57.9\% & 22 & 6 & 72.7\% \\
SPECT Heart   & 83,887,088  & 83,886,215 & 0.0\%  & 50,332,292 & 50,331,748 & 0.0\%  & 23     & 23     & 0.0\%     & 50,332,340 & 50,331,724 & 0.0\%  & 186   & 10   & 94.6\%  & 171   & 161  &     5.8\%  & 24     & 24     & 0.0\%   \\ \bottomrule
\end{tabular}
}
\end{table}

\begin{table}[ht]
\caption{Gate count in the decomposed quantum circuit for QasmSimulator when $r=2$.  P stands for P-PQM, EP~--- for EP-PQM, and S~--- for relative savings.}\label{tbl:sim_decomposed_gates_cnt}
\resizebox{\linewidth}{!}{
\begin{tabular}{lrrrrrrrrrrrrrrr}
\toprule
              & \multicolumn{3}{c}{Quantum Circuit Depth}                                                     & \multicolumn{3}{c}{\qk{cx}}                                                                & \multicolumn{3}{c}{\qk{measure}}                                                  & \multicolumn{3}{c}{\qk{u}}                                                         & \multicolumn{3}{c}{\qk{u3}}                                                                \\
              \cmidrule(r){2-4} \cmidrule(r){5-7}  \cmidrule(r){8-10}  \cmidrule(r){11-13}  \cmidrule(r){14-16}
              & P                            & EP                         & S                          & P                            & EP                         & S                          & P                      & EP                      & S                          & P                       & EP                      & S                          & P                            & EP                         & S                          \\  \midrule
Balance Scale & 20,972,442 & 81,987 & 99.6\% & 12,583,499 & 49,277 & 99.6\% & 21 & 13 & 38.1\% & 145 & 61 & 57.9\% & 12,583,740 & 49,316 & 99.6\%  \\
SPECT Heart & 83,887,094 & 83,886,221 & 0.0\% & 50,332,295 & 50,331,751 & 0.0\% & 23 & 23 & 0.0\% & 171 & 161 & 5.8\% & 50,332,558 & 50,331,766 & 0.0\% \\ \bottomrule
\end{tabular}
}
\end{table}

\end{landscape}
\clearpage
}
\clearpage
\printbibliography

\end{document}